\documentclass[10pt,twocolumn,letterpaper]{article}

\usepackage{framed}
\usepackage{color}
\definecolor{shadecolor}{rgb}{0.92,0.92,0.92}
\usepackage{cvpr}
\usepackage{makecell}
\usepackage{multirow}
\usepackage{times}

\usepackage{caption}

\usepackage{epsfig}
\usepackage{graphicx}
\usepackage{amsmath}
\usepackage{amssymb}
\usepackage{amssymb}
\usepackage{pifont}
\newcommand{\xmark}{\ding{53}}%
\usepackage{times}
\usepackage{epsfig}
\usepackage{graphicx}
\usepackage{amsmath}
\usepackage{amssymb}
 \usepackage{indentfirst}
\usepackage{subfigure}
\usepackage{caption}
\usepackage{graphicx}
\usepackage{authblk}
\makeatletter
\newcommand*\bigcdot{\mathpalette\bigcdot@{.5}}
\newcommand*\bigcdot@[2]{\mathbin{\vcenter{\hbox{\scalebox{#2}{$\m@th#1\bullet$}}}}}
\makeatother

\newcommand{\fref}[1]{Figure.~\ref{#1}}
\newcommand{\sref}[1]{Section.~\ref{#1}}
\newcommand{\tref}[1]{Table.~\ref{#1}}

\makeatletter
\newcommand{\printfnsymbol}[1]{%
  \textsuperscript{\@fnsymbol{#1}}%
}

\usepackage[pagebackref=true,breaklinks=true,colorlinks,bookmarks=false]{hyperref}

\cvprfinalcopy 

\def\cvprPaperID{****} 

\begin{document}
\title{DW-GAN: A Discrete Wavelet Transform GAN for NonHomogeneous Dehazing}

\author[1]{Minghan Fu\thanks{Authors contributed equally} }
\author[1]{Huan Liu\printfnsymbol{1}}
\author[1]{Yankun Yu}
\author[1]{Jun Chen}
\author[2]{Keyan Wang}
\affil[1]{Department of Electrical and Computer Engineering, McMaster University, Hamilton, Canada}
\affil[2]{State Key Laboratory of Integrated Service Networks, Xidian University, Xi'an, China}
\affil[ ]{\textit {\{fum16,liuh127,yuy142,chenjun\}@mcmaster.ca}, \textit {kywang@mail.xidian.edu.cn}}
\affil[ ]{* Authors contributed equally }
\renewcommand\Authands{ and }

\twocolumn[{%
\renewcommand\twocolumn[1][]{#1}
\maketitle
\vspace{-8mm}
\begin{center}
\centering
\includegraphics[width=1.0\textwidth]{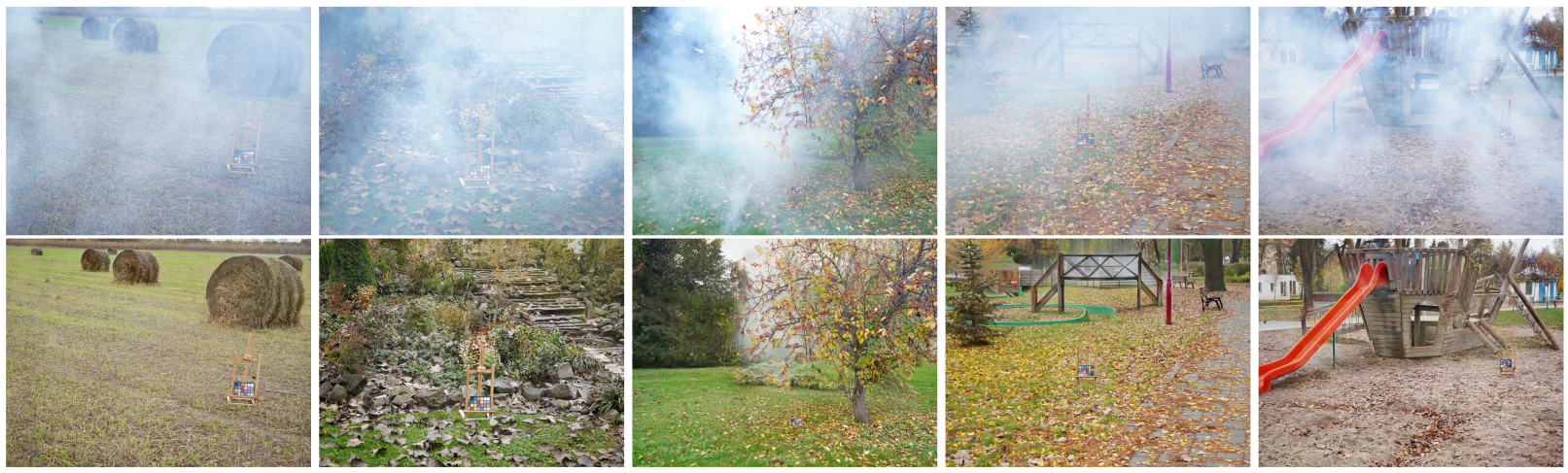}
\captionof{figure}{Our dehazing results on NH-HAZE2 testing set.}
\label{fig:test_nh21}
\end{center}
}]

\pagestyle{empty}  
\thispagestyle{empty} 
\begin{abstract}
Hazy images are often subject to color distortion, blurring, and other visible quality degradation. Some existing CNN-based methods have great performance on removing homogeneous haze, but they are not robust in non-homogeneous case. The reasons are mainly in two folds. Firstly, due to the complicated haze distribution, texture details are easy to be lost during the dehazing process. Secondly, since the training pairs are hard to be collected, training on limited data can easily lead to over-fitting problem. To tackle these two issues, we introduce a novel dehazing network using 2D discrete wavelet transform, namely DW-GAN. Specifically, we propose a two-branch network to deal with the aforementioned problems. By utilizing wavelet transform in DWT branch, our proposed method can retain more high-frequency knowledge in feature maps. In order to prevent over-fitting, ImageNet pre-trained Res2Net is adopted in the knowledge adaptation branch. Owing to the robust feature representations of ImageNet pre-training, the generalization ability of our network is improved dramatically. Finally, a patch-based discriminator is used to reduce artifacts of the restored images. Extensive experimental results demonstrate that the proposed method outperforms the state-of-the-arts quantitatively and qualitatively. The source code is available at \url{https://github.com/liuh127/DW-GAN-Dehazing}.
\end{abstract}

\section{Introduction}
Hazy images are often prone to color distortion, blurring and other visible quality degradation. The varied image degradations often lower the perceptual quality of pictures and devastate numerous intelligent systems, such as tracking \cite{tracking} , satellite remote sensing \cite{remotesesoring1,remotesesoring2} , and object detection \cite{objectdetection1,objectdetection2} . Therefore, image dehazing has attained much attention in the computer vision community.
Many previous dehazing methods are based on the classical atmospheric scattering model \cite{formulation}:
\begin{equation}
I(x)=J(x)t(x)+A(x)(1-t(x))
\end{equation}
where  $I(x)$ denotes the hazy image, $J(x)$ represents the clear image, $A(x)$ is the global atmospheric light, $t(x)$ is the medium transmission and $x$ indicate the pixels. Besides,  $t(x)=$ $e^{-\beta d(x)}$. Where $\beta$ and $d(x)$ are respectively the atmosphere scattering parameter and the scene depth. 

Based on the atmospheric scattering formulation,
some prior-based methods have been proposed \cite{DCP,nonlocal,color,colorline}. These methods estimate atmospheric light
$A(x)$ and the medium transmission map
$t(x)$ by hand-crafted priors, such as dark
channel prior \cite{DCP} and non-local
prior \cite{nonlocal}. However, it is quite hard to accurately estimate $A(x)$ and $t(x)$.
Especially in the non-homogeneous dehazing task, the haze distribution is much more complicated and the haze density is not
strongly correlated to the image depth. Therefore, using prior-based method can result in huge estimation error. Such methods are no longer good choices for non-homogeneous dehazing.
Recent years, with the development of deep learning techniques \cite{lecun2015deep}, many deep learning based dehazing methods \cite{dehazenet, AOD, DCPDN, GCA,FFA} have also been proposed. These methods use convolutional neural networks (CNNs) to extract features and learn the mappings directly between hazy and haze-free image pairs. However, these methods usually require a large number of image pairs during the training process. As the training data becomes 
less, many deep learning based methods are harder to succeed. In addition, the high-frequency components in the clear images, such as edges and fine textures, are often degraded significantly by non-homogeneous haze.  Therefore, restoring clear texture details and sharp edges from hazy images are essential for achieving good perceptual quality. 

In summary, difficulties mainly come from two folds in non-homogeneous dehazing. Firstly, due to the complex haze distribution, texture and color details are easy to be lost during restoration. Secondly, the training image pairs are hard to be collected. Using limited data to train for a robust non-homogeneous dehazing model is quite challenging.
To address the above two problems, we propose a two-branch generative adversarial network. In our first branch, we use the designed wavelet down-sampling modules to replace parts of the convolution layers. By doing this, the number of parameters can be reduced. The lightweight model can have better performance on small training datasets and avoids over-fitting problems caused by model redundancy. In addition, the discrete wavelet transform \cite{dwt} retains the frequency domain information in the images and feature maps. Such information is more conducive to the restoration of texture details.
In our second branch, we employ the pre-trained Res2Net \cite{res2net} as the backbone to extract multilevel features. This pre-trained encoder can
bring substantial prior knowledge for small training datasets \cite{he2019rethinking}. By leveraging the prior knowledge, we can observe significant improvements on small-scale datasets regarding testing accuracy. 
Moreover, we further employ an attention mechanism in our pipeline. Pixel-wise attention module and channel-wise attention module allow the network to focus on the hazy zones and more critical channel information.
Finally, the discriminator is used to introduce an adversarial loss in the training stage. By adopting the adversarial loss, our network is pushed to learn for natural and photo-realistic solutions.

Overall, we summarize our contributions as follows:

1. We propose a two-branch end-to-end trainable GAN to address the non-homogeneous dehazing problem. 

2. We introduce a novel way to embed 2D discrete wavelet transform in our proposed network, aiming at preserving sufficient high-frequency knowledge and restoring clear texture details. In order to perform well in small-scale datasets, we adopt the prior feature knowledge by using ImageNet pre-trained weights as initialization.

3. We show extensive experimental results and comprehensive ablation analysis to illustrate the effectiveness of our proposed method.

\begin{figure*}[h]
	\centering
	{
		\begin{minipage}[t]{\textwidth}
			\centering
			\includegraphics[width=0.92\textwidth]{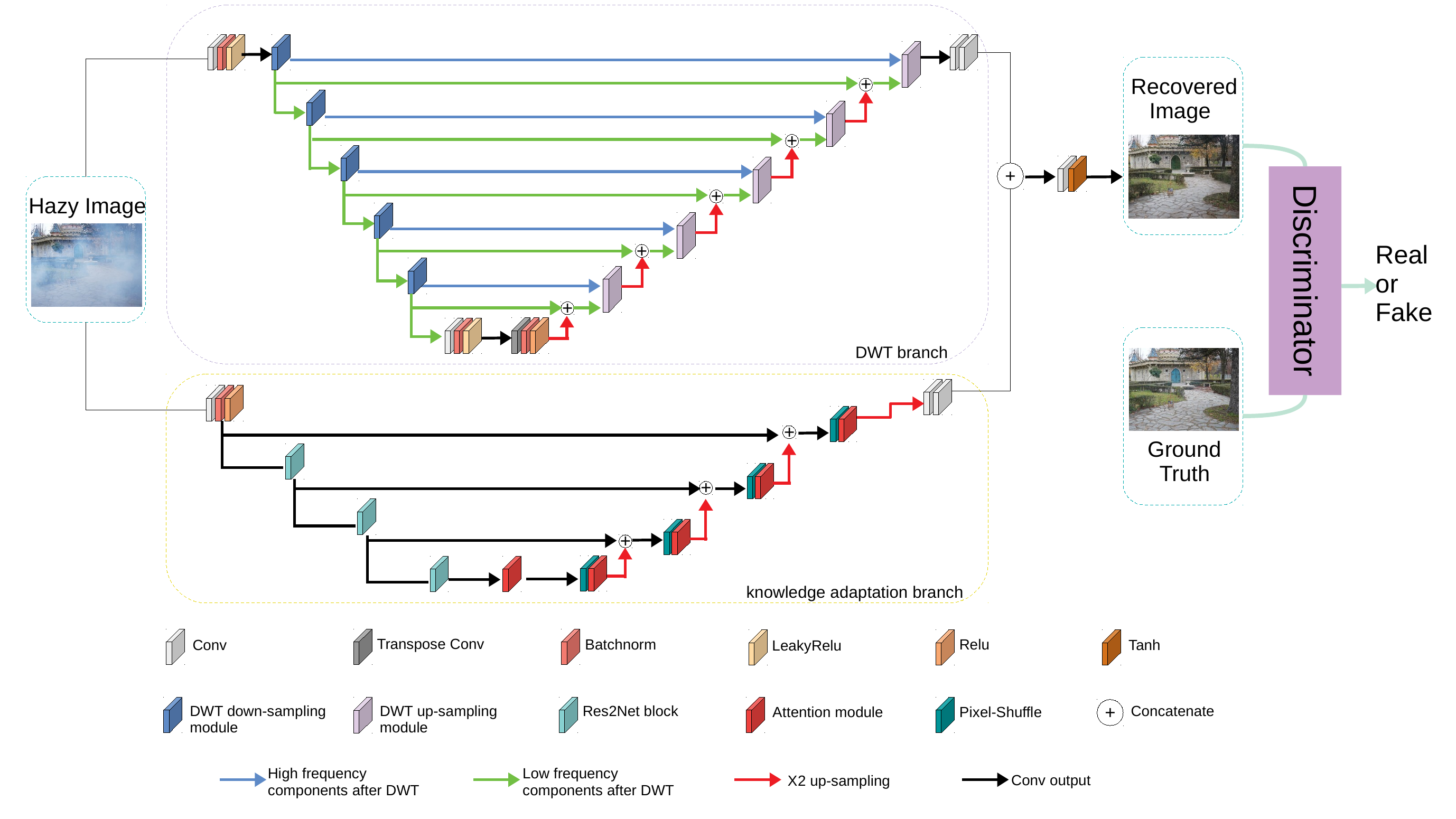}
		\end{minipage}%
	}%
	\caption{The network structure of the proposed method. The generator is a two-branch network, which consists of DWT branch and knowledge adaptation branch. The same color used in the cubic denotes the same operation.}
	\label{pipline}
\end{figure*}
\section{Related Works}
\textbf{Single Image Dehazing.}
Recently, the image dehazing task has attracted intensive attention in the computer vision community. And researchers have proposed many methods for single image dehazing. These methods can be roughly divided into two categories: prior-based methods and learning-based methods.

Prior-based methods utilize prior statistical knowledge, and hand-crafted features for image dehazing. Dark channel prior (DCP) \cite{DCP} is one of the outstanding representatives among prior-based methods. DCP
assumes that hazy images may have extremely low intensities in at least one color channel.
Based on the difference between the brightness and the saturation of hazy image, color attenuation
prior (CAP) \cite{color} creates a linear model to estimate scene depth as strong prior knowledge. 
\cite{nonlocal} hypotheses that colors of haze-free image can be well approximated by a few hundred distinct colors.
Although prior-based methods have achieved good performance in single image dehazing, hand-crafted features and prior knowledge impede these methods to achieve satisfying performance when they are implemented in variable scenes. Recently, with the rapid progress of the deep learning approach \cite{lecun2015deep}, deep learning shows its remarkable ability in solving single image dehazing problem. Some of the deep learning based methods still rely on the atmospheric scattering model. For example,
\cite{dehazenet} proposed DehazeNet as the first end-to-end CNN to learn transmission map. Specifically, it follows the traditional procedure and atmospheric scattering formulation but uses CNN to estimate the transmission map. Similarly, a novel multi-scale neural network (MSCNN) \cite{multidehaze} is then proposed to estimate the transmission map from the hazy image. Different from \cite{dehazenet,multidehaze}, AOD-Net jointly estimates the transmittance and atmospheric light through a lightweight neural network. 
Other than the above-listed methods that depend on the atmospheric scattering model, most recent dehazing methods pay intensive attention to discover a good mapping from hazy image to clear counterpart. 
GFN \cite{gfn} uses white balance, contrast enhancement, and gamma correction methods to pre-process the hazy input. And the dehazed output is then generated by fusing the features of the three derived inputs.  
GCANet \cite{GCA} adopts the smoothed dilation technique into the pipeline for removing the grid artifacts caused by the dilated convolution. FD-GAN \cite{FD-GAN} proposed an end-to-end GAN \cite{gan} with a fusion discriminator. The fusion discriminator integrates the frequency information as additional priors during the training stage. Unlike the \cite{FD-GAN} that discriminates clear image and generated image in the frequency domain, our method focus on directly fusing the high-frequency domain features into the generator.
These deep learning based methods have achieved great performance in homogeneous dehazing task. However, the success is inseparable from the support of large training datasets. We shall empirically demonstrate that they tend to fail drastically in non-homogeneous dehazing with extensive experiments.

\textbf{Frequency Domain Learning.} 
Frequency analysis has always been a powerful tool in image processing. Effective usage of image frequency domain information can greatly improve the performance of the methods in image restoration. Recently, some approaches embedded frequency information into the network structure and exploited the effectiveness of frequency domain information. A wavelet residual network \cite{dwt1} is proposed with the discovery that neural networks can benefit from learning on wavelet subbands. DWSR \cite{dwt2} designed a deep wavelet network that can recover missing details in subbands. MWCNN\cite{dwt3} considered multi-level wavelet transform to enlarge receptive field. AWNet \cite{dai} further utilizes the frequency domain knowledge for image ISP. These methods take advantage of discrete wavelet transform and use it in designing deep learning network architectures.

\textbf{Generative Adversarial Network.}
Generative Adversarial Networks (GANs) \cite{gan} consist of two parts: Generator and Discriminator. They contest with each other via a game theoretic
min-max optimization framework. GANs have achieved great performance in synthesizing realistic images. Many researchers utilize adversarial loss for various low-level vision tasks, such as
image to image translation \cite{imagetoiamge1,imagetoiamge2}, super-resolution \cite{SRGAN}, single image dehazing \cite{hardgan} and image deraining \cite{derain}.

\section{Proposed Method}
In this section, we first describe the overall network architecture (shown in \fref{pipline}) and explain the sense of two-branch designing. Then, we introduce the concept of discrete wavelet transform (DWT) and analyze the benefits of using DWT in our pipeline. In the end, we further demonstrate the loss functions adopted in the training stage.

\begin{figure*}[!t]
	\begin{minipage}{0.5\textwidth}
		\centering
		\includegraphics[width=2in]{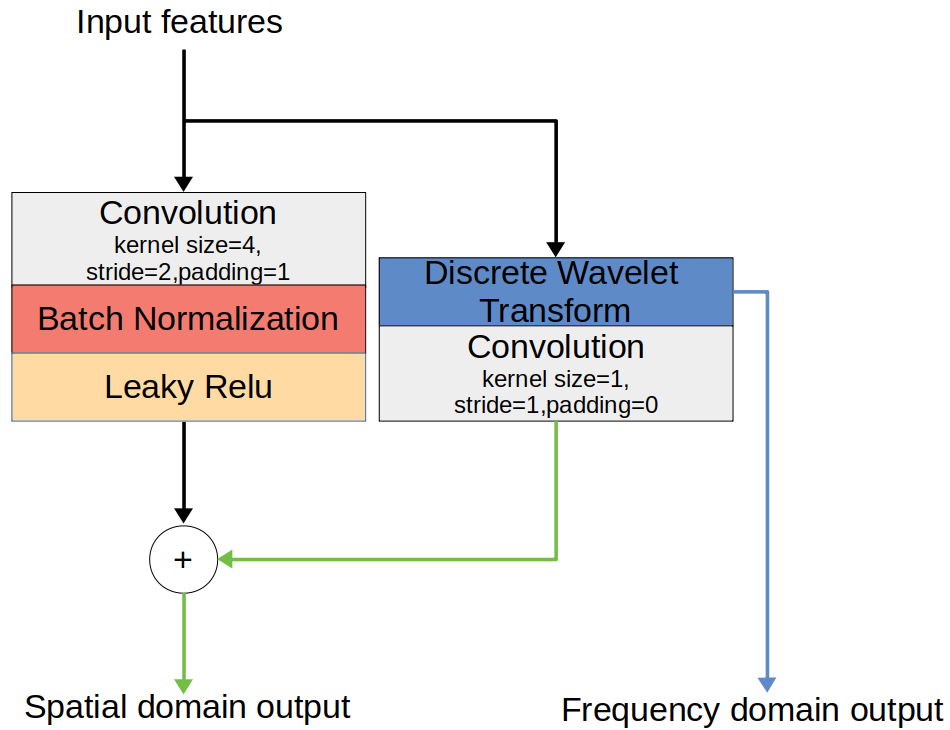}
		\label{fig:a}
	\end{minipage}%
	\begin{minipage}{0.5\textwidth}
		\centering
		\includegraphics[width=2in]{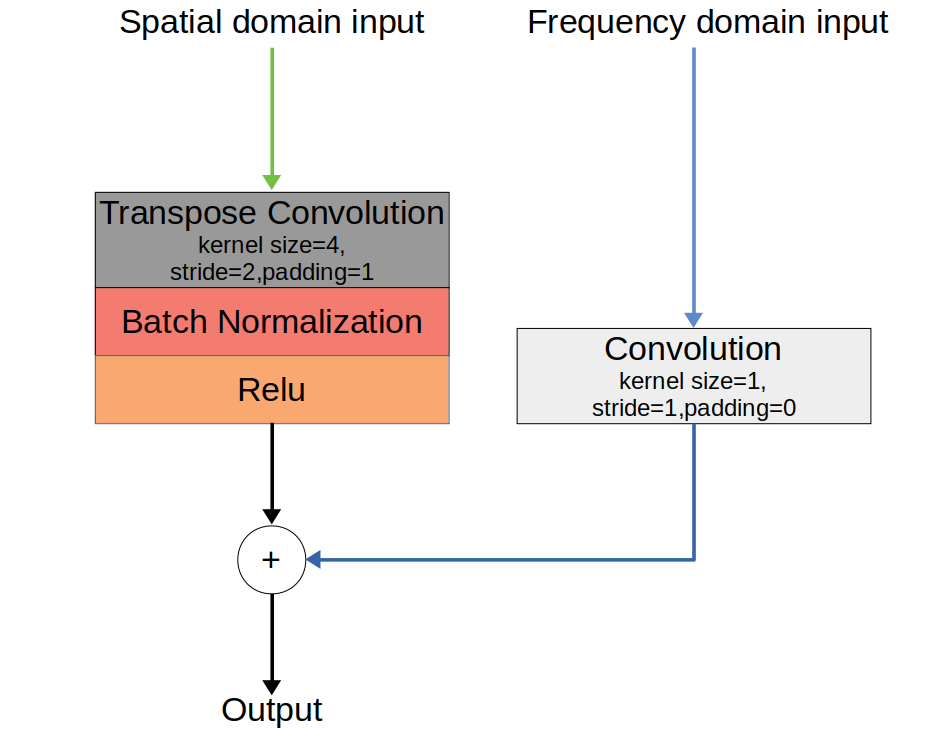}
	\label{fig:b}
	\end{minipage}
	\caption{Left:DWT down-sampling module. Right:DWT up-sampling module.}
	\label{fig:dwt}
\end{figure*}

\subsection{Network Architecture}\label{sec:network}
The two-branch-designed network has been successfully applied in various computer vision tasks \cite{twobranch,multibranch}.
By using this architecture, each network branch can have its own information processing procedures and extract different representations from the same input. In observing that, if we can use such distinct information wisely and make them complement each other by proper fusion strategies, sufficient and comprehensive information from two branches can greatly boost the performance of image dehazing. Based on this idea, we design a two-branch neural network. 

\textbf{DWT Branch.}
Our first branch, \ie, DWT branch (shown in \fref{pipline}) , is designed to directly learn the color mapping from hazy to haze-free images. To achieve this, we follow the U-Net \cite{u-net} to construct our DWT branch. It has an encoder, a decoder, and massive skip connections at each feature scale.
 To meet our requirements for preserving more texture details during dehazing process, we propose to use discrete wavelet transform (DWT) in the feature extraction stage. Since the input feature maps can be decomposed into the low-frequency and high-frequency components by DWT (detailed explanation can be seen in \sref{sec:dwt}), our network can be forced to learn from both high-frequency and low-frequency components. As shown in \fref{fig:dwt}, low-frequency components are concatenated with convolution output as down-sampling features and high-frequency components are added to the DWT up-sampling module by skip connection. By doing this,  our network not only learns abundant information from both spatial domain and frequency domain but also retains favorable image details by high-frequency skip connection.

However, due to the limited data in non-homogeneous dehazing task, it is hard to achieve plausible performance solely relying on DWT branch. Towards better performance on small-scale datasets, we introduce our second branch to utilize additional knowledge further.

\textbf{Knowledge Adaptation Branch.}
Our second branch, \ie, knowledge adaptation branch (shown in \fref{pipline}), focuses on adopting the prior knowledge gained from image classification to the current dehazing task. It leverages the power of transfer learning  \cite{pretrain1,pretrain2} and brings extra information to the small datasets. To achieve this, we use the ImageNet  \cite{imagenet} pre-trained Res2Net \cite{res2net}  as the backbone of our encoder. In the decoder module, we use pixel-shuffle layers for up-sampling, which reduces the computational overload \cite{pixelshffle} and makes the size of feature maps gradually recovered to the original resolution. Besides, inspired by \cite{FFA}, attention blocks are employed after each pixel-shuffle layer to identify the dynamic hazy patterns. In the end, multiple skip connections are added between the encoder and decoder as shown in \fref{pipline}. In this way, our DW-GAN becomes much more robust and has better generalization ability.

Finally, we add a simple $7\times7$ convolution layer as a fusion operation to map the combined features from two branches to clear images. 

\subsection{Discrete Wavelet Transform} \label{sec:dwt}
In $2 \mathrm{D}$ discrete wavelet transform, there are four filters, \ie, low-pass filter $f_{L L}$, and high-pass filters $f_{L H}$,
$f_{H L}$, $f_{H H}$. These filters have fixed parameters with stride 2 convolution operation during the transformation. Thus, by convolving with each filter, images or feature maps can be decomposed into four subbands \ie, $x_{L L}, x_{L H}, x_{H L},$ and $x_{H H} .$ 
 We can express $x_{L L}$ as $\left(f_{L L} \circledast x\right) \downarrow_{2}$, where $\circledast$ represents convolutional operation, $x$ is the input signal and $\downarrow_{2}$ indicates down-sampling by the scale factor of 2. We embed Haar DWT \cite{DWTtheroy} in our proposed method, where
 $f_{L L}=$ $\begin{pmatrix}
\begin{smallmatrix}
1 & 1 \\
1 & 1
\end{smallmatrix}
\end{pmatrix}$, $f_{L H}=$ $\begin{pmatrix}
\begin{smallmatrix}
-1 & -1 \\
1 & 1
\end{smallmatrix}
\end{pmatrix}$, $f_{H L}=$ $\begin{pmatrix}
\begin{smallmatrix}
-1 & 1 \\
-1 & 1
\end{smallmatrix}
\end{pmatrix}$, $f_{H H}=$ $\begin{pmatrix}
\begin{smallmatrix}
1 & -1 \\
-1 & 1
\end{smallmatrix}
\end{pmatrix}$. The $(i, j)$ -th value of $x_{L L}$ after 2D Haar wavelet transform can be defined as:
\begin{equation}
\begin{aligned}
x_{L L}(i, j)=&x(2 i-1,2 j-1)+x(2 i-1,2 j)\\&+x(2 i, 2 j-1)+x(2 i, 2 j)
\end{aligned}
\end{equation}
 The expressions of $x_{L H}, x_{H L},$ and $x_{H H}$ are similar to that of $x_{L L}$. By using DWT, we can obtain the frequency domain knowledge that retains hazy image details, especially from $x_{L H}, x_{H L}$ and $x_{H H}$. However, only using DWT to do image dehazing is not enough. We thus combine these frequency domain operations with convolution so that the network can learn from both spatial and frequency information. Experiment results show the great improvement of using DWT method (see details in \sref{sec:ablation}).


\subsection{Loss Functions}
We denote our dehaze image as $\hat{I}$. $I^{gt}$ and $I^{hazy}$ are respectively the ground truth image and hazy image. The two-branch dehazing network and discriminator represent as $G$ and $D$.

\textbf{Smooth L1 Loss.} $\hat{I}_c(i)$ and $I^{gt}_c(i)$ denote the intensity of the c-th channel of pixel $i$ in the dehazed image and in the ground truth image respectively, and N denotes the total number of pixels. The smooth L1 Loss can be defined as:
\begin{equation}
\mathcal{L}_{\text{smooth-L1}}=\frac{1}{3N} \sum_{i=1}^{N} \sum_{c=1}^{3} \alpha\left(\hat{I}_{c}(i)-I^{gt}_{c}(i)\right)
\end{equation}
where
\begin{equation}
\alpha(e)=\left\{\begin{aligned}
0.5 e^{2}, & \text { if }|e|<1 \\
|e|-0.5, & \text { otherwise }
\end{aligned}\right.
\end{equation}

\textbf{Perceptual Loss.} Besides the pixel-wise supervision, we use the VGG16 \cite{vgg} pre-trained on ImageNet \cite{imagenet} as the loss network to measure perceptual similarity. The loss function is defined as:
\begin{equation}
\mathcal{L}_{\text {per }}=\sum_{j=1}^{3} \frac{1}{C_{j} H_{j} W_{j}}\left\|\phi_{j}(I^{gt})-\phi_{j} (\hat{I})\right\|_2^2
\end{equation}
where $H_j$ , $W_j$ , and $C_j$ denote the height, width, and channel of the feature
map in the j-th layer of the backbone network, $\phi_j$ is the activation of the j-th
layer. $I^{gt}$ and $\hat{I}$ are respectively the ground truth image and our dehazed result.

\textbf{MS-SSIM Loss.} Let $O$ and $G$ denote two windows of common size centered at pixel $i$ in the dehazed image and the haze-free image, respectively. Use a Gaussian filter to $O$ and $G,$ and compute the resulting means $\mu_{O}, \mu_{G},$ standard deviations $\sigma_{O}, \sigma_{G_{1}}$ and covariance $\sigma_{O G} .$ The SSIM for pixel $i$ is defined as:
$$
\begin{aligned}
\text{SSIM}(i) &=\frac{2 \mu_{O} \mu_{C}+C_{1}}{\mu_{O}^{2}+\mu_{G}^{2}+C_{1}} \cdot \frac{2 \sigma_{O G}+C_{2}}{\sigma_{O}^{2}+\sigma_{G}^{2}+C_{2}} \\
&=l(i) \cdot \operatorname{cs}(i)
\end{aligned}
$$
where $l(i)$ represents luminance and $cs(i)$ represents contract and structure measures, $C_{1}, C_{2}$ are two variables to stabilize the division with weak denominator. The MS-SSIM loss is computed using $M$ levels of SSIM. Specifically, we have
$$
\mathcal{L}_{\text{MS-SSIM}}=1-\mathbf{\text{MS-SSIM}}
$$
where
$$
\mathbf{\text{MS-SSIM}}=l_{M}^{\alpha}(i) \cdot \prod_{m=1}^{M} c s_{m}^{\beta_{m}}(i)
$$
with $\alpha$ and $\beta_{m}$ being     default parameters.

\textbf{Adversarial Loss.} The adversarial loss $l_{adv}$ is defined based on the probabilities of the discriminator $D(G(I^{hazy}))$ over all training samples as:
$$
\mathcal{L}_{adv}=\sum_{n=1}^{N}-\log D(G(I^{hazy}))
$$
Here, $D(G(I^{hazy}))$ is the probability of reconstructed image $G(I^{hazy})$ to be a haze-free image.

\textbf{Total Loss.} We combine the smooth L1 Loss, perceptual loss, MS-SSIM loss and adversarial loss together to supervise the training of our dehazing network.
\begin{equation}
\mathcal{L}_{total} = \mathcal{L}_{\text{smooth-L1}} + \alpha \mathcal{L}_{\text{MS-SSIM}} + \beta \mathcal{L}_{per} + \gamma \mathcal{L}_{adv}
\end{equation}
where $\alpha=0.2, \beta=0.001$ and $\gamma=0.005$ are the hyperparameters weighting for each loss functions.

\section{Experiments}
In this section, we firstly describe the datasets that are used for evaluating the effectiveness of our proposed method. Secondly, we introduce our experimental settings \ie, implementation details, and evaluation metrics. Then, we conduct ablation studies to illustrate the benefits of each component in DW-GAN. After that, we compare the performance of our proposed method with the state-of-the-art  qualitatively and quantitatively. Finally, we demonstrate our data pre-processing method and dehazing results in NTIRE2021 NonHomogeneous Dehazing Challenge.

\begin{table*}
	\centering
	
	\setlength{\tabcolsep}{6mm}{
	\begin{tabular}{|c|c|c|c|c|c|c|}
		\hline
		Methods & $l_{1}$ & $l_{p}$ & $L_{SSIM}$ & $l_{adv}$&PSNR&SSIM\\
		\hline
		(1)vanilla DWT branch &$\surd$&\xmark&\xmark&\xmark&18.15&0.7483\\
		(2)knowledge adaptation branch &$\surd$&\xmark&\xmark&\xmark&20.15&0.8156\\
		(3)Two-branch &$\surd$&\xmark&\xmark&\xmark&\underline{21.35}&\underline{0.8273}\\
		(4)Two-branch+DWT &$\surd$&\xmark&\xmark&\xmark&\textbf{21.52}&\textbf{0.8403}\\
		\hline
		(5)Two-branch+DWT&$\surd$&$\surd$&\xmark&\xmark&21.67&0.852\\
		(6)Two-branch+DWT&$\surd$&$\surd$&$\surd$&\xmark&\underline{21.86}&\underline{0.8555}\\
		(7)Two-branch+DWT&$\surd$&$\surd$&$\surd$&$\surd$&\textbf{21.99}&\textbf{0.856}\\
		\hline
	\end{tabular}}
	\caption{Ablation Studies for architectures and loss functions. It can be observed that the model  with all components and supervised by all loss functions performs the best in terms of PSNR and SSIM. }
	\label{tab:ablation}
\end{table*}

\subsection{Datasets}

\textbf{RESIDE Benchmark.} The Indoor Training Set (ITS) of RESIDE\cite{reside} contains 1399 clean images and 13990 hazy images, generated by corresponding clean images with
the medium extinction coefficient $\beta$ chosen uniformly from $[0.6,1.8]$ and the global atmospheric light $A$ chosen uniformly from $[0.7,1.0]$. We use ITS to train our network. For testing, the Synthetic Objective Testing Set (SOTS) is adopted, which contains 500 indoor image pairs.

\textbf{Real-world Dataset.}
We further evaluate our performance on three small-scale real-word datasets: DENSE-HAZE \cite{densehaze}, NH-HAZE \cite{nhhaze1,nhhaze2} and NH-HAZE2 \cite{21report}.
DENSE-HAZE is characterized by dense and homogeneous hazy scenes. It contains 45 training data, 5 validation data and 5 testing data. In our work, we use the official testing data for evaluation and combine the official training set and evaluation set for training our model. NH-HAZE contains 45 training data, 5 validation data and 5 testing data. The haze pattern in this dataset is un-uniformly distributed.  We use 50 training pairs and 5 validation pairs as training set, and use 5 test pairs as testing set. NH-HAZE2 is introduced in NTIRE2021 dehazing challenge. It only contains 25 training data, 5 validation data and 5 testing data. Because the validation and testing set is not public by far, we use image 1-20 as training set and 21-25 as testing set.


\begin{table*}[!t]
	\centering

	\setlength{\tabcolsep}{4.5mm}{
	\begin{tabular}{|c|c|c|c|c|c|c|c|c|}
		\hline
		&\multicolumn{2}{c|}{ITS}&\multicolumn{2}{c|}{NTIRE19}&\multicolumn{2}{c|}{NTIRE20} & \multicolumn{2}{c|}{NTIRE21}  \\
		\cline{2-9}
		&PSNR&SSIM&PSNR&SSIM&PSNR&SSIM&PSNR&SSIM\\
		\hline
		DCP&19.63&0.8823  &11.06&0.4368  &13.28&0.4954  &11.68&0.7090\\
		AOD-Net&19.06&0.8524            &13.21&0.4694    &13.44&0.4136    &13.30&0.4693\\
		GCANet&30.23&0.9814             &12.46&0.4712    &17.49&0.5918    &18.79&0.7729\\
		FFA&\textbf{36.39}&\textbf{0.9886}       &\underline{16.31}&\underline{0.5362}    &18.60&0.6374    &\underline{20.45}&\underline{0.8043}\\
		TDN&34.59&0.9754 &15.50&0.5081    &\underline{20.44}&\underline{0.6683}    &20.23&0.7622\\
		Ours&\underline{35.94}&\underline{0.9860}      &\textbf{16.49}& \textbf{0.5911}    &\textbf{21.51}&\textbf{0.7111}    &\textbf{21.99}&\textbf{0.8560}\\
		\hline
	\end{tabular}}
	\caption{Quantitative comparisons of SOTA methods over SOTS, DENSE-HAZE, NH-HAZE, NH-HAZE2. The best results are in \textbf{bold}, and the second best are with \underline{underline}.}
	\label{tab:sota}
\end{table*}

\begin{table*}
	\centering
	\label{tab:Ablation}
	\setlength{\tabcolsep}{6.5mm}{
	\begin{tabular}{|c|c|c|c|c|c|c|}
		\hline
		 Methods&DCP&AOD-Net&GCANet&FFA&TDN&Ours\\
		\hline
		Inference Time(s)&0.41&0.15&0.65&2.62&0.63&0.48\\
		\hline
	\end{tabular}}
	\caption{Inference time comparisons of SOTA methods on processing one 1600 $\times$ 1200 image.}
	\label{tab:inferece time}
\end{table*}


\subsection{Experimental Settings}
Despite the varied characteristics of each dataset, we adopt the same training strategy for all datasets.
Specifically, we randomly crop patches with a size of 256 $\times$ 256. To augment training data, we implement random rotation (90, 180 or 270 degrees) and random horizontal flip.  
We train DW-NET with the batch size of 16 and utilize Adam optimizer \cite{adam} ($\beta_{1}$=0.9, $\beta_{2}$=0.999). In the training process, a specific decay strategy is used, where the initial learning rate is set to 1e-4 and decays 0.5 times at 3000, 5000, 6000 epoch for total 8000 epochs. The discriminator uses the same optimizer and training strategies. All the experiments are conducted on two NVIDIA 1080Ti GPUs.

\textbf{Quality Measures.} To quantitatively evaluate the performance of our method, we adopt two common metrics: the Peak Signal to Noise Ratio (PSNR) and the Structural Similarity index (SSIM) \cite{SSIM}.

\subsection{Ablation Study} \label{sec:ablation}
Firstly, we conduct comprehensive ablation studies to demonstrate the necessity of each component in our proposed method. 
According to the ablation principle, we construct four different networks to illustrate the importance of each module.
(1) vanilla DWT branch: only use the vanilla DWT branch without DWT down-sampling modules and high-frequency skip connections. (2) knowledge adaptation branch: only use the knowledge adaptation branch to restore hazy images. (3) Two-branch: use two-branch structure, which consists of vanilla DWT branch and knowledge adaptation branch.
(4) Two-branch+DWT: use two-branch structure, where DWT down-sampling and up-sampling modules are embedded into the DWT branch.

\begin{figure*}[!t]
	\begin{minipage}[t]{0.5\linewidth}
		\centering
		\includegraphics[width=3.4in]{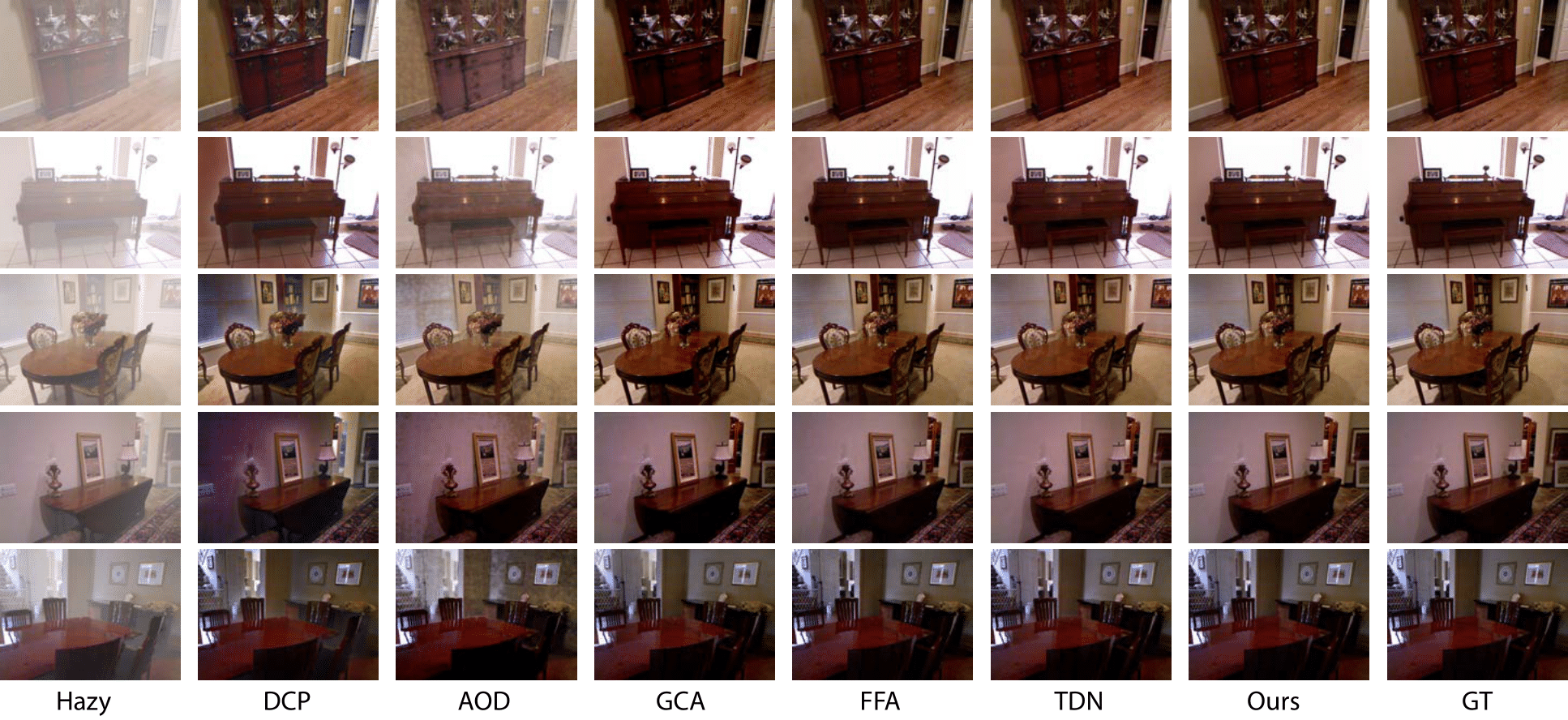} 
	\end{minipage}%
	\begin{minipage}[t]{0.5\linewidth}
		\centering
		\includegraphics[width=3.4in]{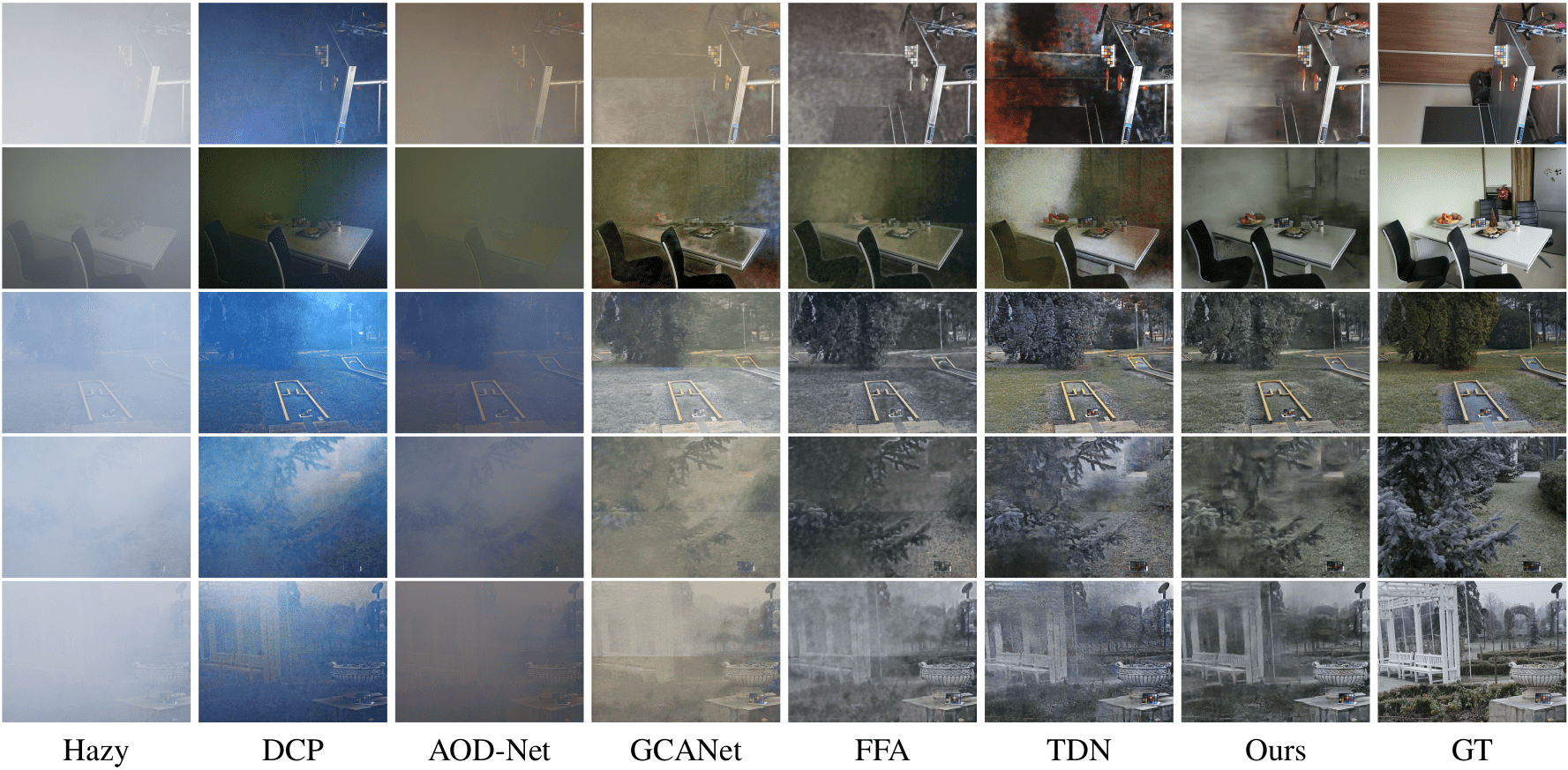}
	\end{minipage}

	\caption{Left: qualitative evaluation examples of RESIDE SOTS indoor testing data. Right: qualitative comparisons of our method with others on DENSE-HAZE dataset.}
	\label{fig:qualitative1}
\end{figure*}

\begin{figure*}[!t]
	\begin{minipage}[t]{0.5\linewidth}
		\centering
		\includegraphics[width=3.4in]{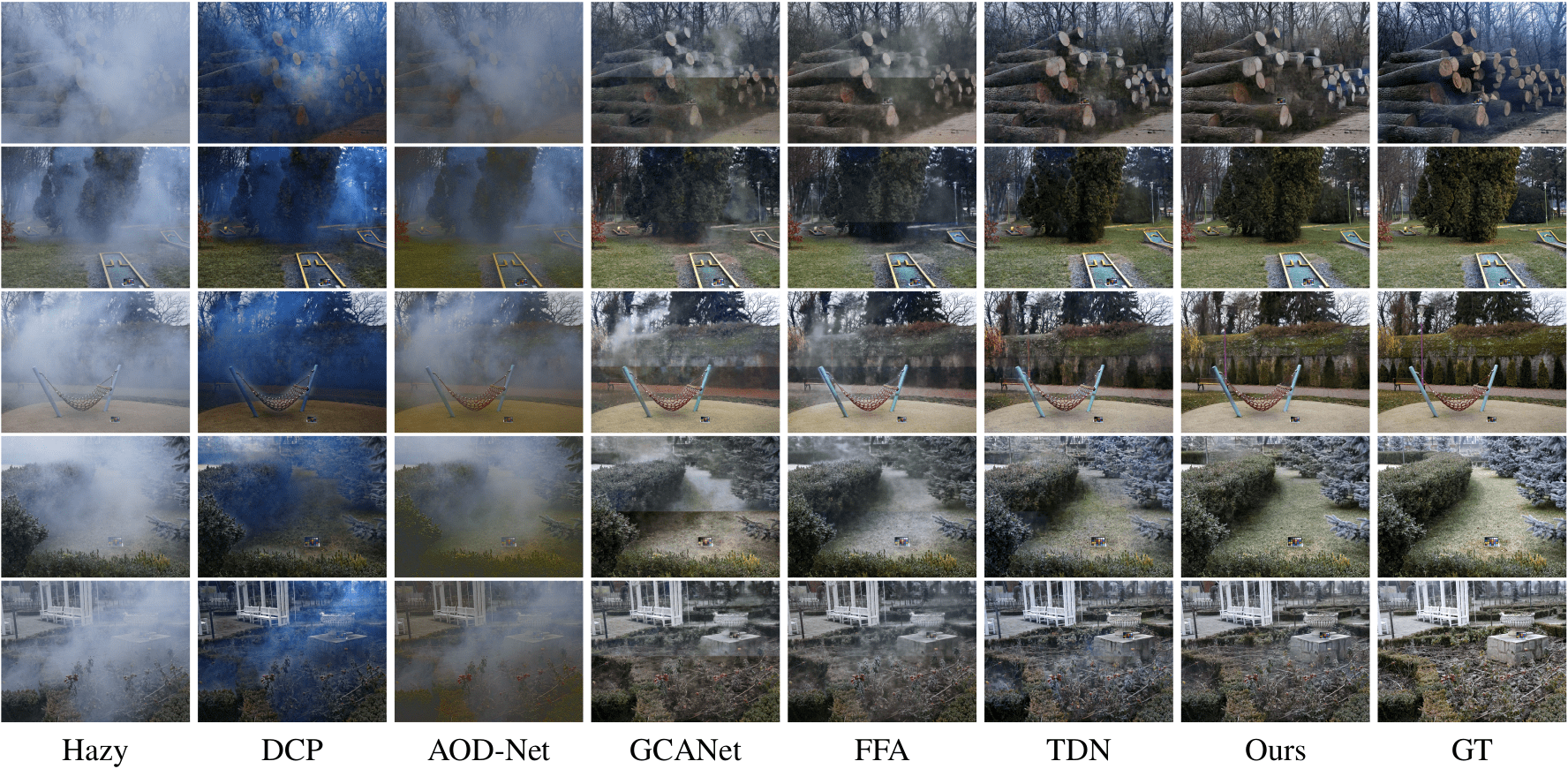} 
	\end{minipage}%
	\begin{minipage}[t]{0.5\linewidth}
		\centering
		\includegraphics[width=3.4in]{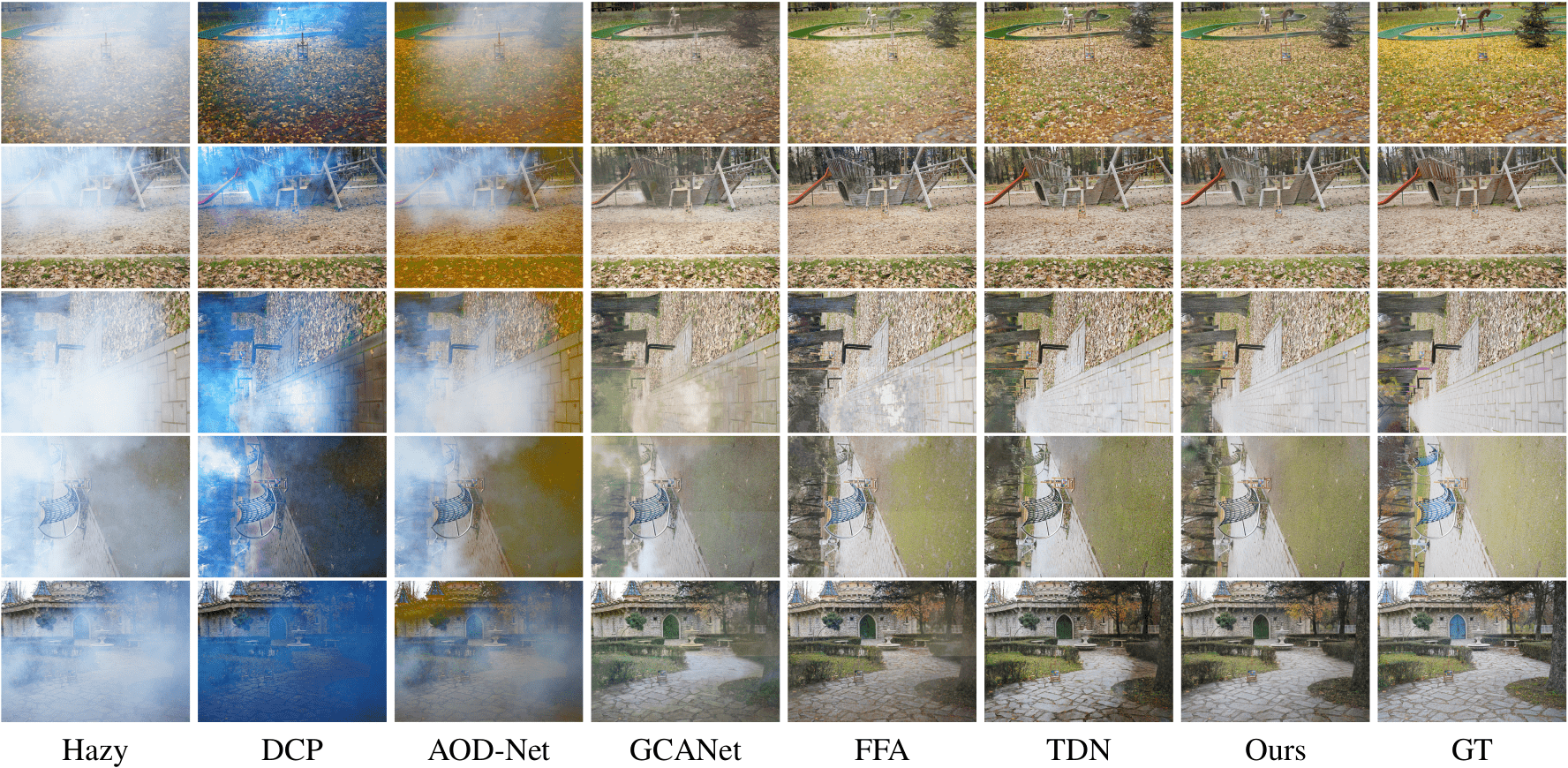}
	\end{minipage}

	\caption{Left: qualitative comparisons of our method with others on NH-HAZE dataset. Right: qualitative comparisons of our method with others on NH-HAZE2 dataset.}
	\label{fig:qualitative2}
\end{figure*}

From the top of \tref{tab:ablation}, we can observe that using two-branch structure can significantly improve our performance in terms of PSNR and SSIM ( by comparing (1), (2) and (3) ). The reason is that two-branch network can not only learn the mapping directly between hazy and haze-free image pairs via vanilla DWT branch but also adapt the pre-learned knowledge to the current task by knowledge adaptation branch. 
To demonstrate the effectiveness of discrete wavelet transform, we compare the cases that the two-branch network adopts DWT or not. By observing the performance of (3) and (4), we can conclude that DWT plays an important role in improving PSNR and SSIM. The increased SSIM also indicates that the frequency domain information is essential for restoring texture details.

Besides, we further illustrate the importance of the loss functions adopted in this work. In observing the fourth to seventh rows of \tref{tab:ablation},
each loss is effective and vital to raising PSNR and SSIM. Smooth L1 loss provides pixel-wise supervision, perceptual loss let the outputs tend to be consistent with ground truth in deep feature space, MS-SSIM loss is employed for minimizing the structural similarity error and GAN loss further improve the output results.
By integrating all the losses on the training stage, our model acquired the best performance (see the last row in \tref{tab:ablation}).

\subsection{Comparisons with State-of-the-art Methods}\label{sec:compare}
We compare the proposed method with state-of-the-art methods on synthetic dataset and real-world datasets. These SOTA methods include DCP \cite{DCP}, AOD-Net \cite{AOD}, GCANet \cite{GCA}, FFA \cite{FFA} and TDN \cite{tdn}. TDN is the winner method in NTIRE 2020 NonHomogeneous Dehazing Challenge.

\textbf{Quantitative Results Comparison.} The experiment results are shown in \tref{tab:sota}. 
For three real-world datasets, our method has outstanding performance and achieves the best in terms of  PSNR and SSIM. 

It is worth noticing that our model has first-class performance on non-homogeneous dehazing and surpassed the second-ranked model by a large margin (1.07dB and 1.54dB higher on NH-HAZE and NH-HAZE2, respectively).
For the synthetic dataset, the performance of our model is slightly lower than that of FFA. The success on large-scale benchmarks often requires heavy network designing. In contrast, we aim to build a suitable model to balance between good mapping capability and over-fitting. But surprisingly, we still perform second best and approach to FFA.

\begin{figure*}[!t]
	\begin{minipage}[t]{0.33\linewidth}
		\centering
		\includegraphics[width=2.5in]{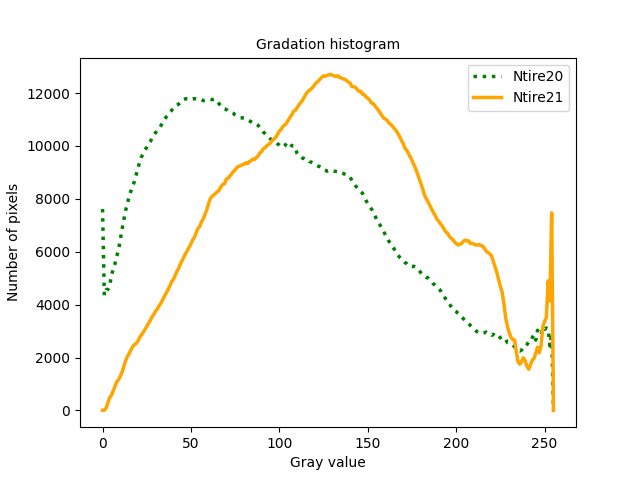} 
	\end{minipage}%
	\begin{minipage}[t]{0.33\linewidth}
		\centering
		\includegraphics[width=2.5in]{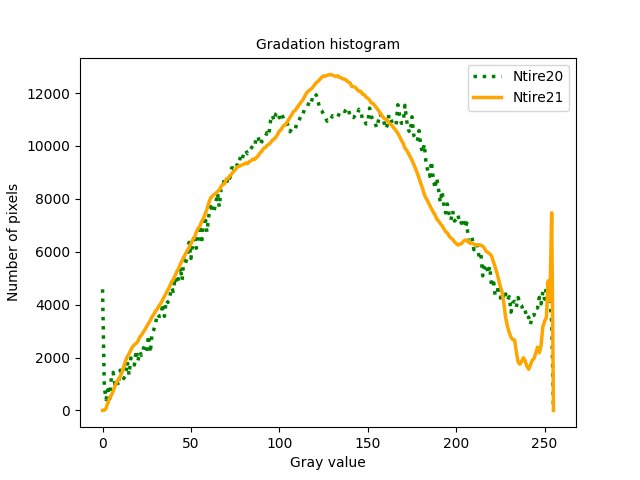}
	\end{minipage}
	\begin{minipage}[t]{0.33\linewidth}
		\centering
		\includegraphics[width=2.3in]{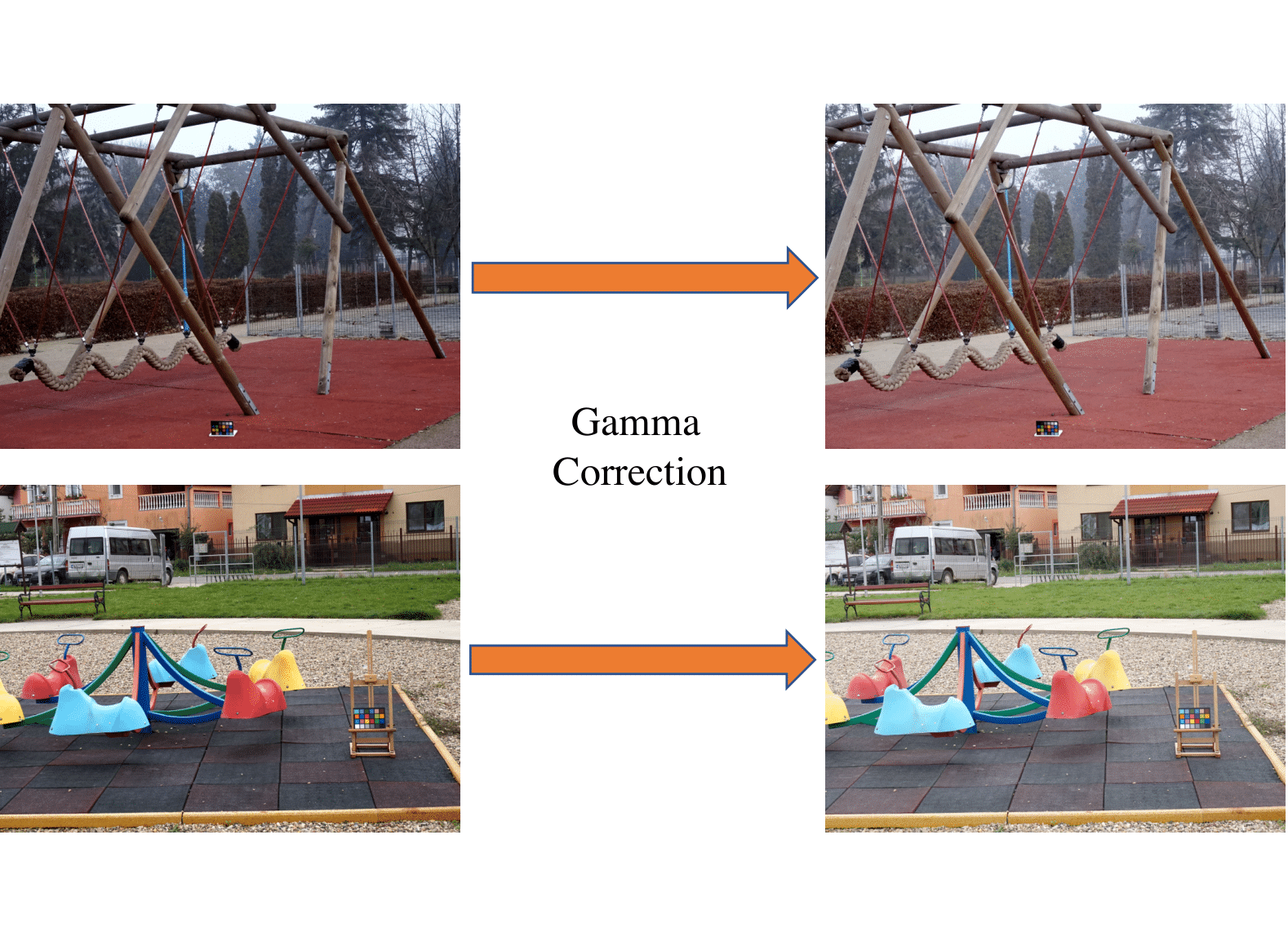}
	\end{minipage}
	\caption{Left:comparison of data distribution without gamma correction. Middle: comparison of data distribution with gamma correction. Right: illustration of gamma corrected clear images in NH-HAZE.}
	\label{fig:validation}
\end{figure*}

\textbf{Qualitative Visual Effect Comparison.}
We show the qualitative results in \fref{fig:qualitative1} and \fref{fig:qualitative2}.
DCP gets much brighter results on ITS test set and bluer results on real-world datasets. The output results of AOD-Net often suffer from severe color distortion and incomplete haze removal on real-world datasets. 
Although GCANet and FFA perform better than the above two methods, they still fail to handle the hazy zones. GCANet tends to generate blurry and color distorted images, and it is unable to remove DENSE-HAZE. Despite the success of  FFA on ITS, it performs comparably bad in the non-homogeneous dehazing task. For example, in NH-HAZE and NH-HAZE2, FFA cannot remove haze effectively and produce unpleasant artifacts. Surprisingly, TDN  shows unsatisfied results in DENSE-HAZE and NH-HAZE2. For example, a considerable color deviation between dehazed images and ground truths can be observed in the DENSE-HAZE dataset. The brightness of dehazed images is much darker, and image details are not restored well on NH-HAZE2. It is worth pointing out that our proposed method performs well on all the datasets, which further reveals the robustness of our model. It can be seen that our dehazed images are visually pleasing and closest to the ground truths.

\textbf{Inference Time Comparison.} We compare the inference time with these SOTA methods for processing one 1600 $\times$ 1200 image by an NVIDIA 1080Ti GPU. As shown in \tref{tab:inferece time}, AOD-Net and DCP take less time to complete dehazing processing. However, these two methods cannot remove haze effectively (details have been discussed in \sref{sec:compare}). It is a decisive fact that our proposed method takes less running time than GCANet, FFA, and TDN. Meanwhile, our approach has better performance both qualitatively and quantitatively.

\subsection{NTIRE2021 Dehazing Challenge}

\textbf{Discussion of Data Pre-processing.}
NTIRE2021 Non-Homogeneous Dehazing Challenge provides only 25 training pairs. To augment training data, we mix image pairs from  NH-HAZE, which consist of 55 non-homogeneous hazy images and clear counterparts. However, images in NH-HAZE and NH-HAZE2 have huge differences in terms of brightness. The visual effect of images in NH-HAZE is much darker, while that of NH-HAZE2 is brighter. To further verify our observation, we quantitatively analyzed the gray-scale distribution of the two datasets (see in \fref{fig:validation} (left)). The average gray value of all haze-free images in NH-HAZE is 102.30 and the variance is 62.42, while in NH-HAZE2, the statistics is 131.45 and 57.45 separately. Due to the difference in brightness, if we simply adopt the model trained with NH-HAZE as extra data to restore hazy images in NH-HAZE2, the average brightness of these dehazed images should lower than 131.45 and higher than 102.3. The inaccurate brightness estimation may result in unsatisfied performance.
In order to reduce the brightness discrepancy between two datasets, we use gamma correction on NH-HAZE. When gamma value is set to 0.65, the average gray value of NH-HAZE is shifted to 133.30 and the variance is changed to 57.78.  With the pre-processing, the gray scale distribution of NH-HAZE is much more similar to that of NH-HAZE2 (shown in \fref{fig:validation} (middle)) and tonal styles of the two datasets are closer.

\textbf{Perform ance on NTIRE2021 Dehazing Challenge.} From the reported  results \cite{21report}, our DW-GAN is among the top performed methods in terms of PSNR and SSIM. To be specific, our dehazed results achieves plausible PSNR (21.08dB) and SSIM (0.8393). To visually demonstrate our performance,  we show the testing results of our DW-GAN in \fref{fig:test_nh21}. It can be observed that our DW-GAN can remove the most of haze and generate visually pleasing results. 
However, we still can find a failure case in \fref{fig:test_nh21} (the third image). Our future work will consider to further exploit a method for addressing those severe hazy areas.

\subsection{Conclusion}
In this paper, we propose a novel generative adversarial network for single image dehazing, namely DW-GAN. DWT-branch directly learns the image mapping from hazy to haze-free images, and leverages the power of discrete wavelet transform in helping the network acquire more frequency domain information. The knowledge adaptation branch exploits the prior knowledge by using pre-trained Res2Net as an encoder. Extra information from the heterogeneous task, \ie, image classification, is introduced to complement the small-scale datasets, which allows our DW-GAN to be much more robust in dealing with limited real-world data. Extensive experimental results illustrate that DW-GAN has great performance in synthetic datasets, real-world scenes with dense haze and non-homogeneous haze. 


{\small
\bibliographystyle{ieee}
\bibliography{egbib}
}

\end{document}